# Unveiling Defect-Mediated Carrier Dynamics in Monolayer Semiconductors by Spatiotemporal Microwave Imaging

Zhaodong Chu[1†], Chun-Yuan Wang[1†], Jiamin Quan[1†], Chenhui Zhang[2], Chao Lei[1], Ali Han[2], Xuejian Ma[1], Hao-Ling Tang[2], Dishan Abeysinghe[1], Matthew Staab[1], Xixiang Zhang[2], Allan H. MacDonald[1], Vincent Tung[2], Xiaoqin Li[1], Chih-Kang Shih[1], Keji Lai[1*]

[1]Department of Physics, University of Texas at Austin, Austin, Texas 78712, USA

[2]Physical Sciences and Engineering Divison, King Abdullah University of Science and Technology, Thuwal 23955-6900, Kingdom of Saudi Arabia.

[†] These authors contributed equally to this work

[*] E-mail: kejilai@physics.utexas.edu

## Abstract

The optoelectronic properties of atomically thin transition-metal dichalcogenides are strongly correlated with the presence of defects in the materials, which are not necessarily detrimental for certain applications. For instance, defects can lead to an enhanced photoconduction, a complicated process involving charge generation and recombination in the time domain and carrier transport in the spatial domain. Here, we report the simultaneous spatial and temporal photoconductivity imaging in two types of $WS_2$ monolayers by laser-illuminated microwave impedance microscopy. The diffusion length and carrier lifetime were directly extracted from the spatial profile and temporal relaxation of microwave signals respectively. Time-resolved experiments indicate that the critical process for photo-excited carriers is the escape of holes from trap states, which prolongs the apparent lifetime of mobile electrons in the conduction band. As a result, counterintuitively, the photoconductivity is stronger in CVD samples than exfoliated monolayers with a lower defect density. Our work reveals the intrinsic time and length scales of electrical response to photo-excitation in van der Waals materials, which is essential for their applications in novel optoelectronic devices.

## I. Introduction

Semiconducting transition-metal dichalcogenides (TMDs) $MX_2$ (M = Mo, W; X = S, Se) exhibit remarkable electrical and optical properties (1-5), such as strong light-matter interaction in the infrared/visible range (6), good carrier mobility (7), and high photoresponsivity (8, 9). As in any materials, however, imperfections like structural defects are inevitable in TMD monolayers (10). Depending on the targeted application, the effect of impurities can be either detrimental or beneficial. For electronic devices, crystalline disorders are usually unwanted because of the adverse effect on carrier mobility (11, 12). For optoelectronic devices, however, the situation is more complicated. Certain defects are favorable for processes like single-photon emitting (13-16), which are important in quantum-emitters applications. Similar considerations also apply for photovoltaics (17) and photodetectors (18, 19), where the dynamics of photo-generated carriers play a key role. Photoconduction is a complex process involving the generation and recombination of electrons and holes, as well as the carrier diffusion from the illumination spot and transport under the electric field. While defects usually suppress carrier mobility, they may also elongate the lifetime of photo-carriers. Their impact on the intrinsic photoconductivity is thus nontrivial. It is imperative to systematically investigate the spatiotemporal evolution of photo-excited carriers in various TMD monolayers with different levels of defects.

The conventional method to spatially resolve photoconduction is scanning photocurrent microscopy (SPCM) (20-23), which measures the transport across electrical contacts as a focused laser beam scans over the sample. The temporal response of this technique is limited by the carrier transit between electrodes and charge transfer at the metal-semiconductor junction, rather than the intrinsic photoresponse time of the material. The spatial resolution of SPCM is also diffraction-limited. In this work, we report the first spatiotemporal study of photo-excited carriers in two types of $WS_2$ monolayers by laser-illuminated microwave impedance microscopy (iMIM), a unique optical-pump-electrical-probe technique with 50-nm spatial resolution and 10-ns temporal resolution. Surprisingly, the photoconductivity is in general stronger in the more defective regions and samples. Such counterintuitive observations are reconciled by our simultaneous spatial and temporal studies of the photo-carrier dynamics in both chemically grown and mechanically exfoliated monolayers. Our results provide fundamental knowledge on the spatiotemporal evolution of charge carriers in 2D TMDs, paving the way for their applications in novel optoelectronic devices.

## II. Experimental Results

We begin our investigation on a chemical-vapor deposited (CVD) WS$_2$ monolayer (see Methods), hereafter denoted as Sample A, which presumably has a high level of structural defects (24, 25). The sample is coated by 30 nm atomic-layer deposited (ALD) Al$_2$O$_3$ to avoid direct tip-sample contact. Conventional characterization data of the monolayer using atomic-force microscopy (AFM), micro-Raman, and PL mapping are included in Supplementary Information S1. Microwave imaging is carried out using the configuration in Fig. 1*A*, where the sample is illuminated by a fiber-coupled green laser ($h\nu$ = 2.4 eV) through the double-side-polished sapphire substrate. The microwave signal at the frequency $f \sim 1$ GHz is delivered to a sharp tip. The real and imaginary parts of the tip-sample admittance are detected by the electronics to form iMIM-Re and iMIM-Im images, respectively (26). Unlike the previous work where the tip was aligned with the laser spot during the sample scan (27-29), there are two sets of piezo stages in the current configuration, one moving the focused laser beam to the desired location and the other carrying the tip to scan over the sample and image the diffusion of photo-excited carriers (Fig. 1*B*). Fig. 1*C* shows the optical image of Sample A with the laser spot in the middle. The iMIM images with and without illumination are displayed in Fig. 1*D*. The line profiles across the optical image on a CCD camera and iMIM images at $P_c = 1.6 \times 10^6$ mW/cm$^2$ are plotted in Fig. 1*E*, where $P_c$ is the laser intensity at the center of the illuminated spot. While the laser shows a Gaussian profile $e^{-r^2/w^2}$ with a width of $w \sim 1.5$ µm, the spatial spread of the iMIM signals clearly exceeds the illumination spot due to carrier diffusion. Interestingly, a slight increase of the iMIM signals is observed at the sample edge, which will be discussed later.

Since the pertinent dimension in this experiment is much smaller than the free-space electromagnetic wavelength at 1 GHz, the iMIM is essentially a near-field impedance probe and the tip-sample interaction is quasi-electrostatic in nature (26 – 29). Quantitative analysis of the iMIM images can be obtained by finite-element modeling (Supplementary Information S2). The localized iMIM response from the region underneath the tip with a dimension comparable to its diameter (Supplementary Information S3) allows us to convert the signals to the local sheet conductance $\sigma_{2D}$ shown in Fig. 1*F*. Under the steady-state condition, the spatial distribution of carrier density $n(r)$ follows the diffusion equation (30, 31)

$$n(r) - L^2 \Delta n(r) = \eta \frac{P_c \tau}{h\nu} e^{-r^2/w^2} \tag{1}$$

where $L = \sqrt{D\tau}$ is the diffusion length, $\Delta = \nabla^2$ the Laplace operator, $\eta$ the incident photon-to-current conversion efficiency (IPCE), $\tau$ the carrier lifetime, and $D$ the diffusion coefficient. The analytical solution to Eq. (1) for a Gaussian laser profile is

$$n(r) \propto \int_{-\infty}^{\infty} K_0(r'/L) e^{-(r-r')^2/w^2} dr' \qquad (2)$$

where $K_0$ is the modified Bessel function of the second kind. In the iMIM experiment, the measured 2D sheet conductance can be expressed by

$$\sigma_{2D} = nq\mu \propto P_c \cdot \tau \cdot \mu \qquad (3)$$

where $q$ is the elemental charge and $\mu$ the carrier mobility. It is clear that the photoconductivity scales with the carrier lifetime and mobility, both affected by the presence of defects. Assuming that $\mu$ is independent of $n$ within the range of our experiment, we can fit the measured $\sigma_{2D}$ profiles to Eq. (2). The extracted diffusion length $L = 2.0 \pm 0.3$ µm is independent of the laser power, suggesting that bimolecular charge recombination plays a minor role under the steady-state condition here, as is neglected in Eq. (1). We have also confirmed that, as expected, $L$ does not depend on the spot size of the laser beam (Supplementary Information S4).

To probe the temporal dynamics of photo-excited carriers, we modify the iMIM setup for time-resolved measurements (32). As illustrated in Fig. 2*A*, the laser output is modulated by an electric-optical modulator (EOM), which is driven by a 200 kHz square wave from a function generator. The iMIM signal reaches the steady state within the time of 2.5 µs for each on/off cycle. The low-*f* filter at the iMIM output stage is also replaced by a high-*f* amplifier. The microwave signals are synchronized to the EOM and measured by a high-speed oscilloscope. We will only present the iMIM-Im data since the signal is roughly proportional to $\sigma_{2D}$ within our measurement range (Supplementary Information S2). Fig. 2*B* shows the time-resolved iMIM (tr-iMIM) data averaged over 8192 cycles at two locations of Sample A for $P_c = 2.4 \times 10^6$ mW/cm$^2$. We note that the steady-state photoconductivity is higher at the edge than that at the center. The apparent rise time $t_{\text{rise}} \sim 10$ ns is limited by the temporal resolution of the instrument. On the other hand, the much longer fall time is clearly resolved by our setup. In Figs. 2*C* and 2*D* (10 times more averaging than that in Fig. 2*B*), the tr-iMIM signals are plotted in the semi-log scale and fitted to bi-exponential functions $A_1 e^{-t/\tau_1} + A_2 e^{-t/\tau_2}$. Two relaxation time constants, $\tau_1$ around 100 ns and $\tau_2$ around 1 µs, are observed in both curves. The amplitudes of the two terms are such that $A_1/A_2 \sim$ 2 at the center of the WS$_2$ flake and ~ 0.45 at the edge. Fig. 2*E* shows the measurement at the same location as Fig. 2*C* but with a higher optical excitation $P_c = 2.2 \times 10^7$ mW/cm$^2$. A rapid drop of iMIM signal with $t_{drop}$ ~ 10 ns, again limited by the temporal resolution, is observed in the beginning of the decay process. It should be noted that finite temporal resolution of iMIM simply means that any processes faster than 10 ns will appear as sudden jumps or rises at that time scale. It does not, however, affect the steady-state signal at $t = 0$ in Figs. 2*C* – 2*E*. The multiple time scale indicates that several recombination mechanisms are relevant for photoconductivity in Sample A.

The iMIM data of Sample A are to be contrasted against that of the mechanically exfoliated flake, hereafter denoted as Sample B. In this work, we have carefully isolated WS$_2$ monolayers from bulk crystals, annealed them under optimal conditions (see Methods), and protected the surface with ~ 10 nm hexagonal boron nitride (h-BN), as shown in the inset of Fig. 3*A*. The much narrower PL spectrum (Supplementary Information S5) of Sample B than Sample A indicates that the former is cleaner (33), consistent with the general view that exfoliated TMD monolayers are usually less defective than the CVD grown monolayers (34, 35). Surprisingly, during our measurement on over 10 TMD samples, we have systematically observed that the photoconductivity in exfoliated monolayers is always lower than that in CVD ones. Because of the weak iMIM response, a rather blunt tip has to be used to enhance signals in the time-resolved experiments and a high-resolution diffusion mapping is difficult. As seen in Fig. 3*A*, at the lowest laser intensity ~ $5 \times 10^5$ mW/cm$^2$, the tr-iMIM data show the exponential relaxation with τ ~ 100 ns, which is comparable to τ$_1$ in Sample A. As the laser power increases, however, a resolution-limited rapid process develops in the tr-iMIM signals. As plotted in Fig. 3*B*, the amplitude of the fast process increases sharply as a function of the laser intensity, whereas the slow process shows a much weaker power dependence. At $P_c = 3 \times 10^6$ mW/cm$^2$, the fast process has completely dominated the tr-iMIM data. In contrast, for the CVD-grown Sample A, the carrier relaxation still follows the slow dynamics up to this power level and the steady-state iMIM-Im signals roughly scale with the laser intensity, as plotted in Fig. 3*C*.

### III. Discussions

The spatiotemporally resolved iMIM experiments reveal rich and quantitative information of WS$_2$ monolayers. For Sample A, taking the weighted average of <τ$_{center}$> ~ 400 ns and $L$ ~ 2 μm, we obtain a diffusion coefficient of $D \sim 0.1$ cm$^2$/s. Using the Einstein relation $D = \mu\,(k_BT/q)$, where $k_BT$ is the thermal energy, one can show that $\mu \sim 4$ cm$^2$/V·s, well within the range of literature values of electron mobility (1 ~ 10 cm$^2$/V·s) in CVD-grown WS$_2$ monolayers (36). Based on our transport data (Supplementary Information S6) and previous iMIM results (27), it is clear that the majority of photo-excited carriers are electrons in the conduction band. Moreover, by substituting the above parameters into Eq. (1), it is straightforward to show that $\eta \sim 0.1\%$ (Supplementary Information S7), again consistent with previous reports of IPCE in monolayer TMDs (37, 38). Finally, using Eq. (3), where $\sigma_{2D}$ is obtained from the iMIM data and mobility from the Einstein relation, we can also convert the photoconductivity at various excitation power to the carrier density, as seen in Supplementary Information S8.

As above-gap photons always excite electron-hole pairs in semiconductors, the predominance of electron transport in photoconduction implies that holes are localized by defects. The presence of structural defects in WS$_2$ is well documented in the literature (10, 34, 39, 40). Fig. 4*A* displays scanning transmission electron microscopy (STEM) images of Sample A, with three types of defects (S, S-S, W vacancies) clearly resolved. Using first-principles density functional theory (DFT), we can calculate the energy band structures and density of states of these defect configurations. As shown in Fig. 4*B*, there exist in-gap states that are able to induce Shockley-Read-Hall (SRH) (41, 42) trap-assisted recombination of mobile carriers. In particular, the mid-gap states are most effective to annihilate both electrons and holes, whereas the band-tail states are either electron or hole traps, depending on their proximity to the conduction or valence band, respectively. Since as-grown WS$_2$ monolayers are weakly n-doped, it is reasonable to assume that the electron traps are filled. Within the field of view of our STEM data, 5 S-S and W vacancies, which lead to trap states near the valence band, are identified. A rough estimate of the defect density is therefore $5 \times 10^{12}$ cm$^{-2}$.

As illustrated in Fig. 4*C*, the iMIM results allow us to systematically analyze the various processes in the photoresponse of monolayer WS$_2$, which is rather different from traditional semiconductors. Upon photon absorption, electrons and holes are rapidly generated ($\tau_g = 10 \sim 100$ fs $\ll 1$ ns) and quickly thermalized to the band extrema ($\tau_{th} = 10 \sim 100$ fs $\ll 1$ ns) (43). After that, electron-hole pairs can recombine through various processes. For radiative recombination, mobile electrons and holes only exist for a short time ($< 1$ ps) before forming charge-neutral excitons (not depicted in the schematic). Note that the exciton lifetime $\tau_r = 1 \sim 100$ ps (44 – 46) is not relevant in this experiment as the typical electric field at the iMIM tip cannot dissociate the tightly bound excitons in monolayer TMDs (Supplementary Information S2). For non-radiative recombination, mid-gap states annihilate photo-generated carriers within a lifetime $\tau_n < 1$ ns, as measured by optical and THz pump-probe experiments (47, 48). Band-tail states, on the other hand, trap holes in WS$_2$ at a fast rate of $\tau_{tr}^{-1}$ and slowly release them at a rate of $\tau_{esc}^{-1}$, after which the escaped holes can recombine with electrons within $\tau_r$ or $\tau_n$. Since $\tau_{esc}$ is much longer than $\tau_{tr}$, the process effectively prolongs the apparent lifetime of electrons in the conduction band (49), which dominates the steady-state photoconductivity at low photon dose. For Sample A, the observation of two relaxation time constants ($\tau_1 \sim 100$ ns, $\tau_2 \sim 1$ μs) indicates the presence of a distribution of trap states in the sample. During the CVD growth, it is likely that the outer edge of the flake is more defective than the center, corresponding to a longer average lifetime measured in Fig. 2*D*. According to Eq. (3), if the dependence of carrier mobility on defect density is relatively modest, the photoconductivity is, counterintuitively, higher in the more defective regions and samples. When the optical excitation increases above a threshold, e.g., $P_c \sim 2 \times 10^7$ mW/cm$^2$ for Sample A, the corresponding carrier density of $4 \times 10^{12}$ cm$^{-2}$ is comparable to the defect density estimated from the STEM data. Consequently, trap states are saturated and the excess carriers are annihilated through much faster recombination processes, consistent with the sudden drop in the tr-iMIM data in Fig. 2*E*. The onset $P_c$ for Sample B, however, is only $5 \times 10^5$ mW/cm$^2$, indicative of much fewer traps in this sample.

As a concluding remark, our findings of the spatiotemporal dynamics in TMDs are important for optoelectronic devices. While ultrafast photoresponse can be detected by pump-probe experiments (44 – 48), the peak intensity of pulsed laser is usually too high for practical photodetectors. In contrast, the response time of previously demonstrated TMD photodetectors is usually on the order of milli-seconds (9, 49 – 51) or even seconds (22, 52), which is clearly limited by the transport through source/drain contact electrodes. The sub-μs carrier lifetime due to trap states can be inferred from photocurrent measurements (49), which requires device modeling and indirect analysis of the data. In that sense, photoresponse directly probed by the tr-iMIM in the ns – μs time scale is not well covered by conventional measurements. A non-contact microwave probe similar to our iMIM setup would take advantage of the sub-μs intrinsic time scale and operate in a much faster speed. Furthermore, while much effort has been made to minimize defects in TMD samples for higher carrier mobility, a moderate amount of defects could actually enhance the photoconductivity, which is favorable for certain device applications.

**IV.     Conclusion**

In summary, we report simultaneous studies of nanoscale spatial distribution and time evolution of photo-generated carriers in $WS_2$ monolayers by near-field microwave imaging. The diffusion length and carrier lifetime in the CVD-grown sample are quantitatively measured and analyzed by the diffusion equation. The calculated carrier mobility using the Einstein relation is consistent with the transport result and that in the literature. Time-resolved measurement shows that the 0.1 ~ 1 μs dynamics are driven by the slow release of holes from the trap states. Beyond a threshold optical excitation, which is higher in CVD than exfoliated monolayers, the traps are filled and excess carriers are then annihilated by much faster recombination channels. By resolving the spatiotemporal dynamics of carriers, our work represents an important step toward controlling and manipulating carriers in advanced nano-optoelectronic devices based on van der Waals materials.

## Methods

**Sample preparation.** For the CVD-grown sample on double-side polished sapphire substrate, solid precursors, including $WO_3$ (300 mg, Sigma Aldrich Co., 99.995%) and S powders (Sigma Aldrich Co., 99.5%), were used for the synthesis. The growth of monolayer $WS_2$ was carried out in a 1-inch furnace with a heating rate of 25 °C/min and growth time of 5 min at 1050 °C. During the growth, the carrying gas of 70 sccm Ar was mixed with 7 sccm of $H_2$, and the gate valve was regulated to keep the pressure at 11.5 mTorr. The S powders were initially heated up to 130 °C for 10 min by a heating tape before the furnace reached the growth temperature. After the growth, the samples were naturally cooled down to ambient temperature under the same $H_2$/Ar flow.

The exfoliated $WS_2$ sample was prepared by a dry viscoelastic stamping method. Both h-BN and monolayer $WS_2$ flakes (bulk crystal from 2Dsemiconductors Inc.) were mechanically exfoliated from bulk crystals onto a polydimethylsiloxane sheet. Prior to the transfer, we thoroughly cleaned the sapphire substrates with acetone and IPA in an ultrasonic bath and subsequent annealing at 800 °C for 10 hours. The bottom $WS_2$ monolayer was first transferred onto the sapphire substrate. The

~ 10 nm hBN was then transferred on top of the WS$_2$ monolayer. After each transfer process, the sample was annealed under ultrahigh vacuum (around $10^{-5}$ mbar) at 300 °C for 6 hours.

**iMIM and tr-iMIM setup.** The microwave imaging experiment was performed in a customized chamber (ST-500, Janis Research Co.) with positioners and scanners (AttoCube Systems AG). The PtIr probe tip (12PtIr400A, Rocky Mountain Nanotechnology LLC.) was mounted on the xyz-scanner for tip scans. For the diffusion mapping experiment, the CW laser intensity was modulated by an optical chopper with a frequency of 1 kHz to improve the signal-to-noise ratio. For the tr-iMIM experiment, the low-$f$ amplifier (54 dB, 20 kHz) after the microwave mixer was replaced by a fast amplifier (28 dB, 350 MHz. SR445A, Stanford Research Systems Inc.). The diode laser was modulated by an electric-optical modulator (M350-160-01 EOM, Conoptics Inc.) with a power supply of 8 ns rise/fall time. The EOM was driven by a 200 kHz square wave from a function generator (DG5071, RIGOL Technologies USA Inc.) with < 4 ns rise/fall time. The tr-iMIM signals were measured by a 600MHz oscilloscope (DS6062, RIGOL Technologies USA Inc.) with 5 GSa/s sampling rate. The time resolution of our setup is about 10 ns.

**STEM preparation and characterization.** The WS$_2$ monolayers were coated by polymethylmethacrylate (PMMA) polymer and the PMMA/WS$_2$/sapphire sample was annealed at 120 °C in air for 10 min. The sample was then put into the diluted HF solution for 5 min to separate the PMMA/WS$_2$ film from the sapphire surface. The 400 Mesh Nickel TEM grid was used to pick up the film. The PMMA was then removed by acetone for more than 1 hour. The TEM grid with WS$_2$ crystals was finally baked in vacuum (< $10^{-6}$ torr) at 350 °C for overnight to remove the PMMA residue. The STEM measurement was conducted at 80 kV using a ThermoFisher USA (former FEI Co) Titan Themis Z (40-300kV) TEM equipped with a double Cs (spherical aberration) corrector, a high brightness electron gun (x-FEG) and Fischione STEM detector.

**First-Principles DFT Calculations.** DFT calculations were performed using the Vienna Ab initio Simulation Package (VASP), in which Generalized gradient approximations (GGA) of Perdew-Burke-Ernzerhof (PBE) have been adopted for exchange-correlation potential. In the model, a supercell is composed of $5 \times 5$ WS$_2$ unit cells and a 20 Å thick vacuum. The lattice constant of a unit cell is 3.19 Å. $5 \times 5$ supercells were used for the calculation of energy band structures and density of states. Four types of calculations were performed: pristine monolayer WS$_2$ and three types of defects (W, S or S-S) with defects density $10^{13}$ cm$^{-2}$. During the self-consistent (SC) calculations, the global break condition for the electronic SC-loop is set to be $10^{-7}$ eV and the cutoff energy for the plane wave basis set is 400 eV, and k-point sampling is $8 \times 8 \times 1$ with a Gaussian smearing method, in which the smearing width is 0.05 eV.

## Acknowledgements

The research at UT-Austin, performed collaboratively between K.L., C.K.S., X. L, and A.H.M., was primarily supported by the National Science Foundation through the Center for Dynamics and Control of Materials: an NSF MRSEC under Cooperative Agreement no. DMR-1720595. K.L., C.K.S., and X.L. were also supported by NSF EFMA-1542747. The iMIM instrumentation was supported by the U.S. Army Research Laboratory and the U.S. Army Research Office under grant nos. W911NF-16-1-0276 and W911NF-17-1-0190. K.L. and Z.C. acknowledge the support from Welch Foundation grant F-1814. X.L. acknowledges the support from Welch Foundation grant F-1662. J.Q. acknowledges the support from the Department of Energy, Basic Energy Science grant DE-SC0019398 and a catalysis grant from the University of Texas at Austin. C.Z and X.X.Z. acknowledge the financial support from King Abdullah University of Science and Technology (KAUST), Office of Sponsored Research (OSR) under Award No. CRF-2016-2996-CRG5. V.T. gratefully acknowledges the research award from the Doctoral New Investigator Award from ACS Petroleum Fund (ACS PRF 54717-DNI10, V.T.), Part of the characterization and synthesis of 2D TMDs in this work were performed as User Proposals (#5067 and #5424) at the Molecular Foundry, Lawrence Berkeley National Lab, supported by the Office of Basic Energy Sciences, of the U.S. Department of Energy under Contract No. DE-AC02-05CH11231

## Author contributions

Z.C. and K.L. conceived the project. Z.C. and X.M. performed the iMIM measurements. C.W. grew the CVD samples and measured the PL image. J.Q., D.A., and M.S. prepared and characterized the exfoliated samples. C.Z. and A.H. carried out the STEM measurement. C.Z. and H.T. performed the transport measurement. C.L. conducted the DFT calculations. Z.C. and K.L. performed the data analysis and drafted the manuscript. All authors approved the final version of the manuscript.

## Competing financial interests

The authors declare no competing financial interests.

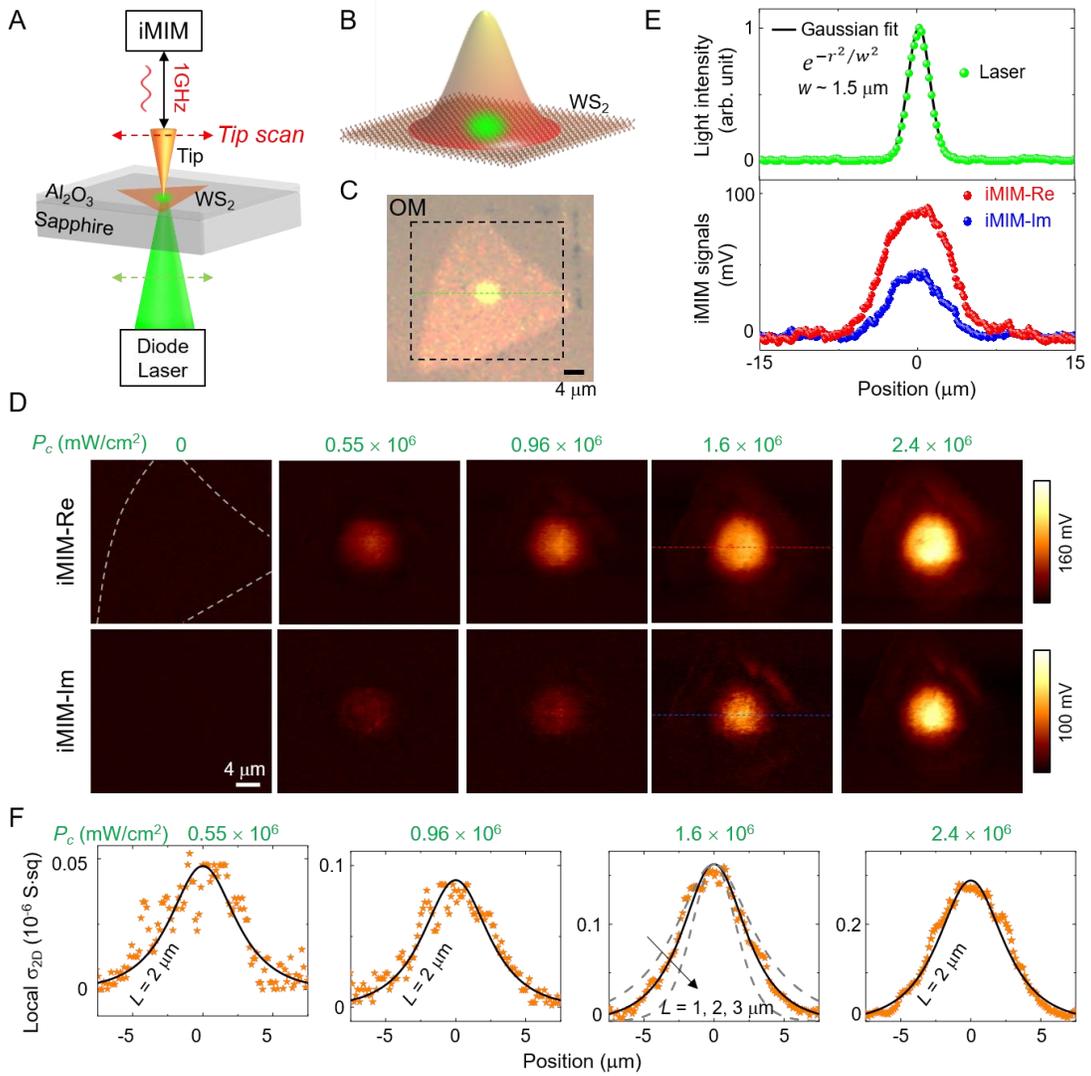

**Figure 1. Diffusion mapping of photo-excited charge carriers**. (*A*) Schematic diagram of the iMIM setup with bottom illumination. Both the tip and the focused laser spot can scan with respect to the sample. The monolayer WS$_2$ flake grown on double-sided sapphire and coated by ALD Al$_2$O$_3$ is also shown in the schematic. (*B*) Illustration of the carrier diffusion from the illumination spot. (*C*) Optical reflection image of Sample A with the laser spot in the middle. (*D*) iMIM images inside the dashed box in *C* at different laser power labeled by the intensity at the center $P_c$. The dashed lines show the contour of the flake. (*E*) Line profiles across the CCD (upper panel) and iMIM images (lower panel) for $P_c = 1.6 \times 10^6$ mW/cm$^2$. The solid line is a Gaussian fit to the laser profile with $w = 1.5$ μm. (*F*) Measured photoconductivity (orange stars) profiles at various laser intensity. The solid lines show the numerical fits with a diffusion length $L = 2$ μm. Dashed lines in the panel with $P_c = 1.6 \times 10^6$ mW/cm$^2$ correspond to $L = 1$ and 3 μm.

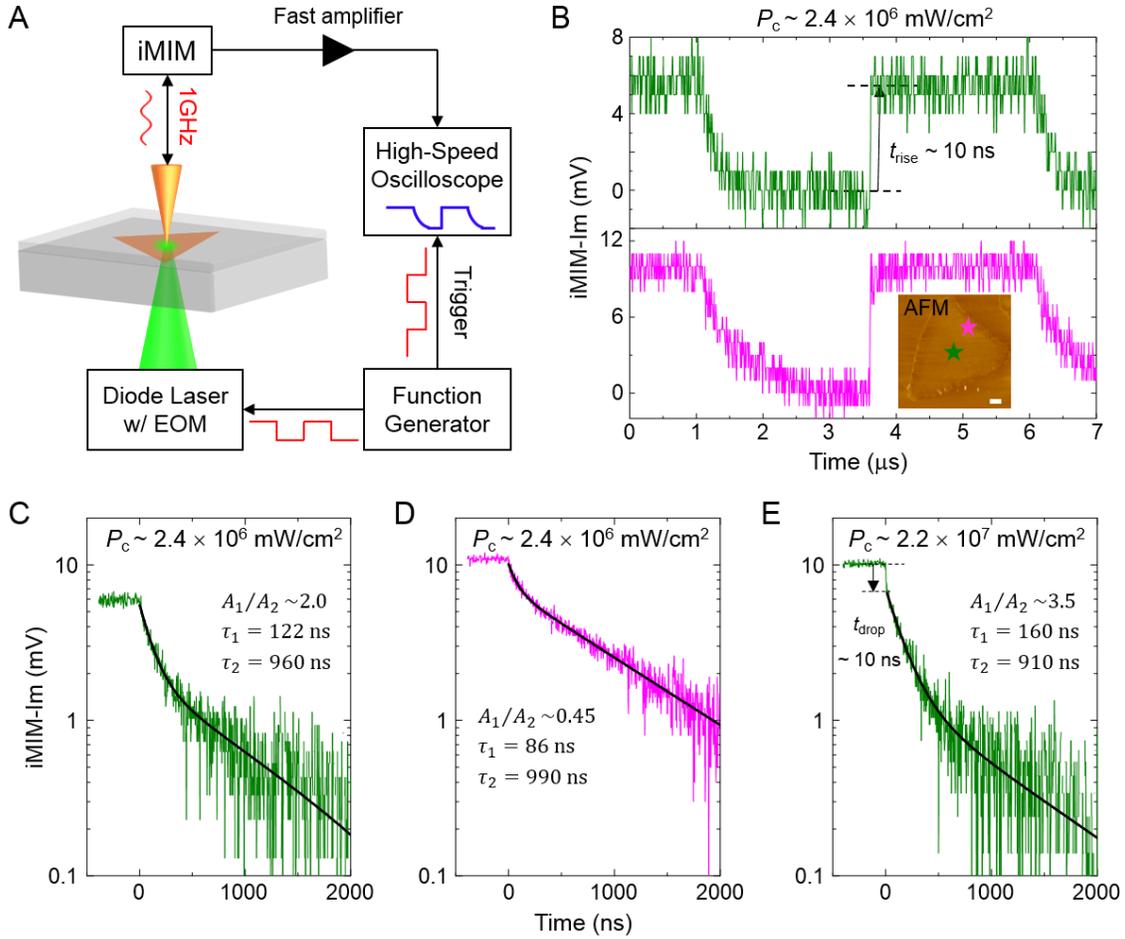

**Figure 2. Time-resolved iMIM and carrier lifetime measurements**. (*A*) Schematic of the tr-iMIM setup, in which the sample is illuminated by a diode laser with its intensity modulated by an EOM. The EOM is driven by a function generator, which also triggers the high-speed oscilloscope that averages the iMIM signals. (*B*) Time-resolved iMIM-Im signals (averaged over 8192 cycles) at the center (green) and edge (magenta) of Sample A. The inset shows the AFM image of the flake, on which the two measured spots are labeled. The scale bar is 4 μm. The rise time of ~ 10 ns, limited by the temporal resolution of the setup, is indicated in the plot. (*C*) Relaxation of tr-iMIM signals, averaged over 10 measurements in *B*, at the center and (*D*) at the edge of the flake. The solid lines are bi-exponential fits to the data with two relaxation time constants. (*E*) Same as *C* but at a higher laser intensity. A rapid drop of the signal with $t_{drop}$ ~ 10 ns, again limited by the temporal resolution, is indicated in the plot.

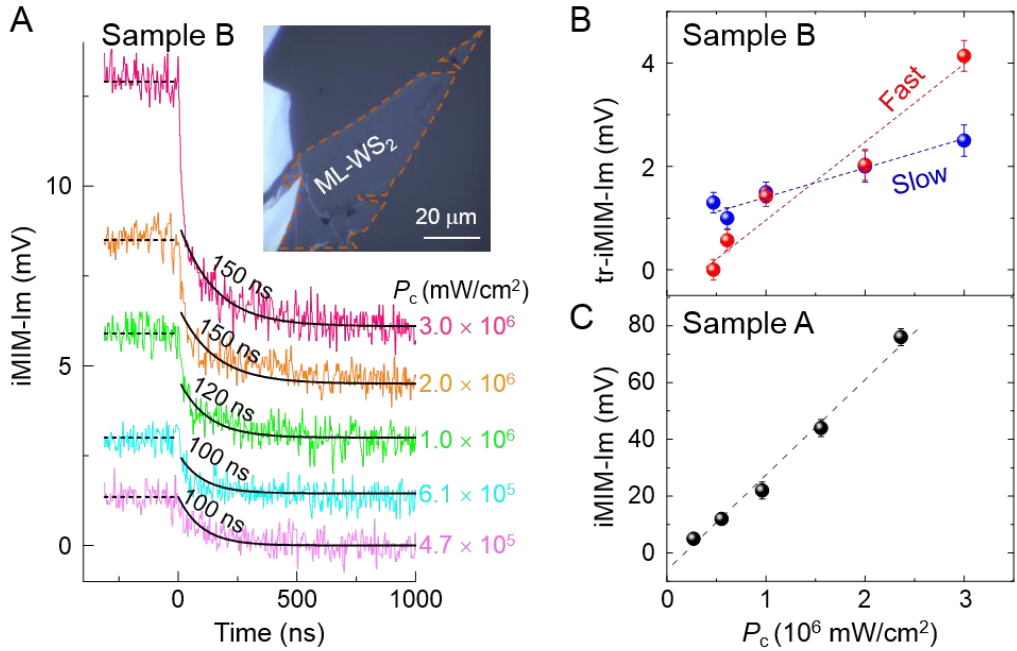

**Figure 3. Time-resolved iMIM results on exfoliated WS$_2$**. (*A*) Time-resolved data on Sample B (optical image in the inset). Except for the lowest laser intensity of $4.7 \times 10^5$ mW/cm$^2$, the tr-iMIM curves all show a sudden drop faster than our temporal resolution, followed by an exponential decay with time constants of 100 ~ 150 ns. (*B*) Amplitudes of the fast (red) and slow (blue) processes in the tr-iMIM data as a function of laser power. The dashed lines are guides to the eyes. (*C*) Steady-state iMIM-Im signals at the center of the laser spot as a function of $P_c$, taken from Figure 1*D*. The dashed line is a linear fit to the data.

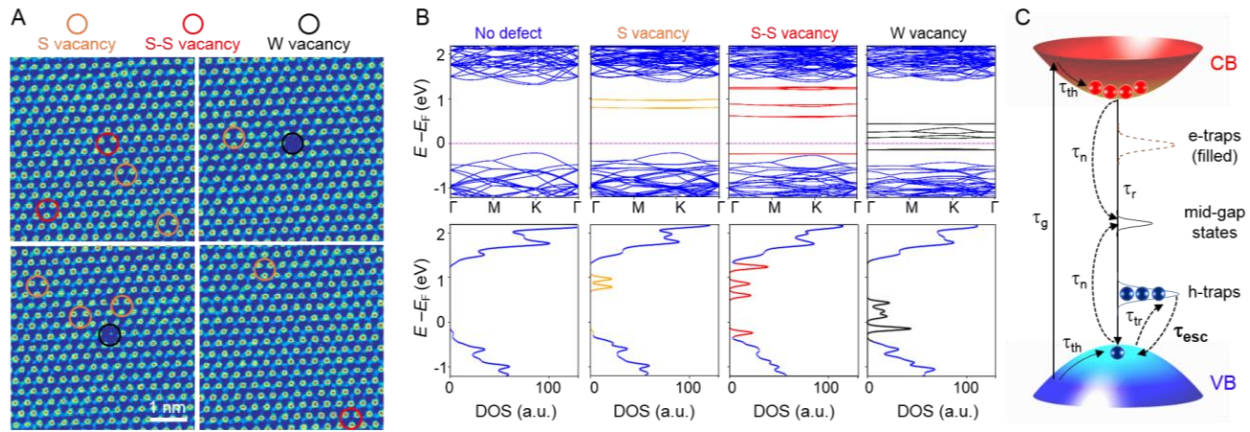

**Figure 4. STEM, DFT, and physical model of the spatiotemporal dynamics**. (*A*) STEM images of the CVD-grown $WS_2$ sample, showing the presence of atomic defects such as S (orange), S-S (red), and W (black) vacancies. (*B*) DFT calculations of the energy band structures and density of states for three types of common atomic defects in $WS_2$ monolayers. (*C*) Schematic diagram of the band structure with mid-gap and band-tail states, as well as multiple processes for charge generation and recombination. $\tau_g$, $\tau_{th}$, $\tau_r$, $\tau_n$, $\tau_{tr}$, and $\tau_{esc}$ represent the time constant for carrier generation, thermalization, radiative recombination, non-radiative recombination, trapping, and escaping from traps, respectively. It is assumed that the electron-traps are filled in n-type $WS_2$ samples, while the hole-traps are dominant in the photoconductivity response.

# Supplementary Information

# Unveiling Defect-Mediated Carrier Dynamics in Monolayer Semiconductors by Spatiotemporal Microwave Imaging

Zhaodong Chu[1†], Chun-Yuan Wang[1†], Jiamin Quan[1†], Chenhui Zhang[2], Chao Lei[1], Ali Han[2], Xuejian Ma[1], Hao-Ling Tang[2], Dishan Abeysinghe[1], Matthew Staab[1], Xixiang Zhang[2], Allan H. MacDonald[1], Vincent Tung[2], Xiaoqin Li[1], Chih-Kang Shih[1], Keji Lai[1*]

[1]Department of Physics, University of Texas at Austin, Austin, Texas 78712, USA

[2]Physical Sciences and Engineering Divison, King Abdullah University of Science and Technology, Thuwal 23955-6900, Kingdom of Saudi Arabia.

[†] These authors contributed equally to this work

* E-mail: kejilai@physics.utexas.edu

## S1. Basic characterization of Sample A.

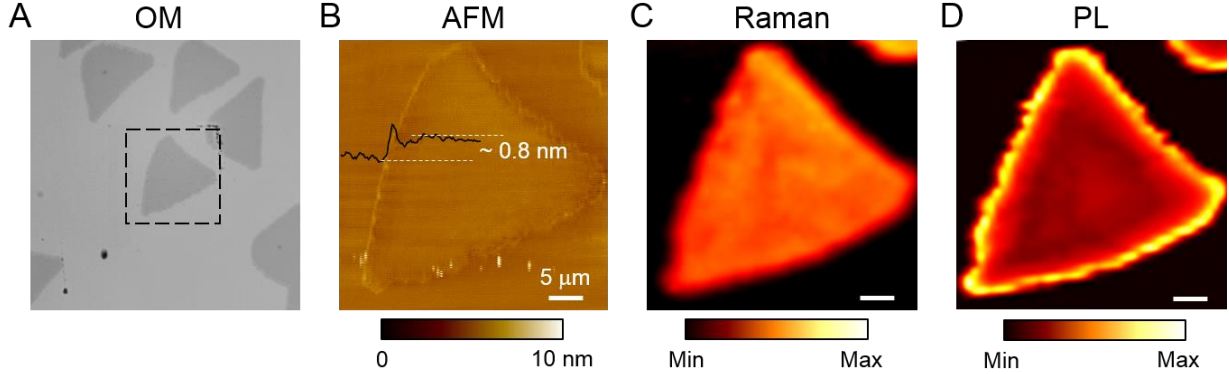

Figure S1. (*A*) Optical image of the CVD-grown monolayer WS$_2$ flakes. (*B* – *D*) AFM, Raman, and PL mapping of the flake inside the dashed box in *A*, respectively. The AFM line profile is averaged over 10 lines, showing a height (~ 0.8 nm) corresponding to the monolayer thickness. The Raman map shows the intensity of the major $E_{2g}^1$ peak in monolayer WS$_2$. The PL map is integrated from 600 to 650 nm. All scale bars are 5 μm.

Characterizations of the CVD-grown WS$_2$ flakes using traditional methods are shown in Fig. S1. The height of the sample is ~ 0.8 nm, consistent with the monolayer thickness. The strong Raman peak at ~ 360 cm$^{-1}$ and PL peak at ~ 640 nm also indicate that the sample is a WS$_2$ monolayer [S1, S2].

## S2. Finite-element analysis of the iMIM signals.

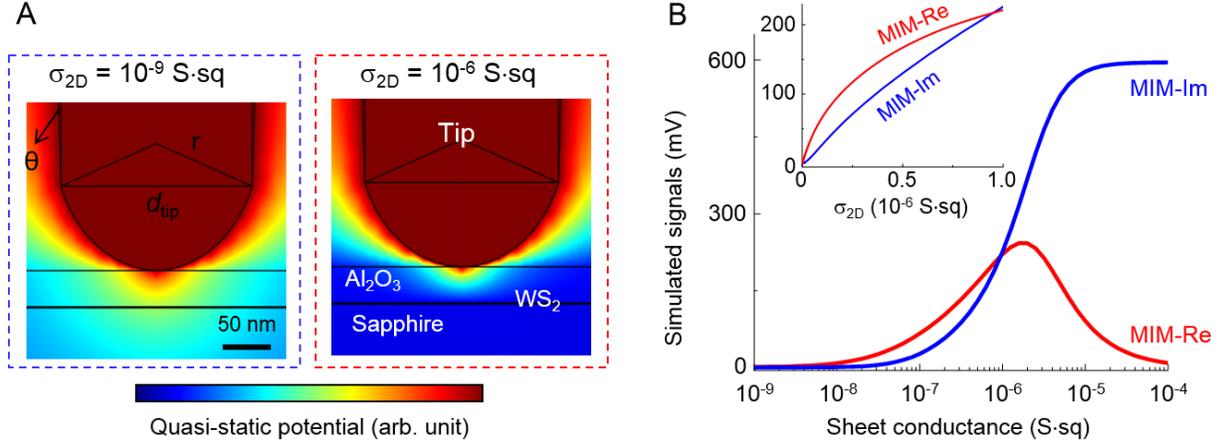

Figure S2. (*A*) Tip-sample geometry and quasi-static distributions when the sheet conductance of WS$_2$ $\sigma_{2D}$ equals to $10^{-9}$ (left) and $10^{-6}$ (right) S·sq. (*B*) Simulated MIM signals as a function of $\sigma_{2D}$ in semi-log scale to cover the entire range of the response curve. The inset shows the MIM response curve up to $10^{-6}$ S·sq in the linear scale.

Finite-element analysis (FEA) [S3] of the iMIM response to 2D sheet conductance is shown in Fig. S2. For the tip-sample geometry in Fig. S2A, the tip shape is defined by $d_{\text{tip}} = 200$ nm, $r = 100\sqrt{3}$ nm, and $\theta = 2°$. The sample parameters are as follows. The monolayer WS$_2$ is 1 nm in thickness, with a dielectric constant $\varepsilon_{\text{WS}_2} = 3.5$ [S4]; the capping layer Al$_2$O$_3$ is 30 nm thick with $\varepsilon_c = 9$ [S5]; the sapphire substrate is 200 μm thick with $\varepsilon_s = 9$ [S5]. The full simulated iMIM response curve as a function of the sheet conductance of WS$_2$ $\sigma_{2D}$ is shown in Fig. S2B. The iMIM-Im signal saturates beyond $\sigma_{2D} \sim 10^{-5}$ S·sq, whereas the iMIM-Re signal displays a maximum at $\sigma_{2D} \sim 2 \times 10^{-6}$ S·sq. The relevant $\sigma_{2D}$ ($< 10^{-6}$ S·sq) in this work, however, is relatively small. Within this narrow range, the iMIM-Im signal is roughly proportional to $\sigma_{2D}$, as seen in the inset of Fig. S2B. We can therefore approximate the tr-iMIM-Im signals as photoconductivity in Figures 2 and 3 in the main text.

With an input power of 10 μW and quarter-wave impedance-match section [S3], the GHz voltage at the tip is ~ 0.1 V. Using a tip radius ~ 100 nm and permittivity of the capping layer ~ 10, we can estimate that the typical electric field at the sample is around $10^5$ V/m, much smaller than the dipolar field of $10^8$ V/m in TMD excitons [S6]. As a result, the microwave field cannot dissociate the tightly bound charge-neutral excitons in WS$_2$, which do not contribute to the iMIM signals.

## S3. Spatial resolution of the iMIM.

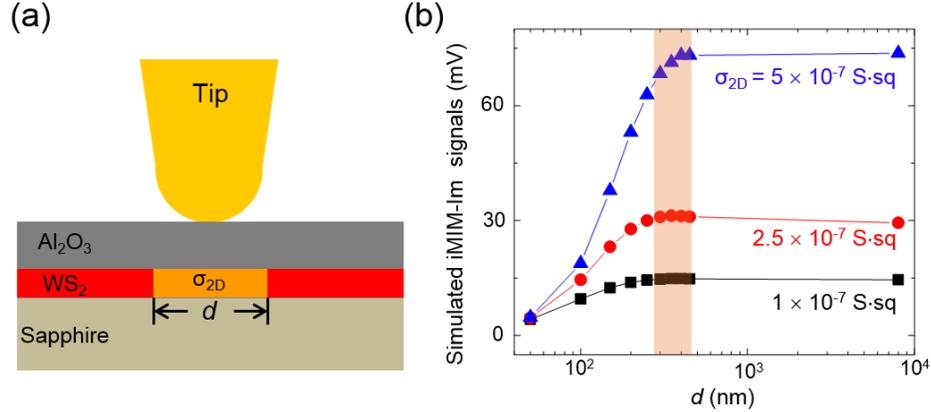

Figure S3. (**A**) Tip-sample configuration for the analysis of spatial resolution. (**B**) Simulated iMIM-Im signals as a function of the width of the $WS_2$ disk underneath the tip.

Strictly speaking, the FEA simulation performed in Fig. S2 does not follow the actual situation in the experiment, where the local conductivity varies continuously in the diffusion map. Fortunately, the iMIM signals come mostly from the region underneath the tip with a spatial dimension comparable to the tip diameter. Such a localized response of the microwave microscope is demonstrated as follows. Fig. S3A shows the modeling configuration, where only a small disk of $WS_2$ (diameter of $d$) has a 2D conductance of $\sigma_{2D}$. For simplicity, the sheet conductance of the rest of the $WS_2$ flake is kept at 0. As plotted in Fig. S3B, the iMIM signals saturate after $d$ is increased beyond ~ 300 nm, confirming the local response and the validity of our interpretation of the iMIM spatial maps.

## S4. Diffusion analysis with a small laser spot.

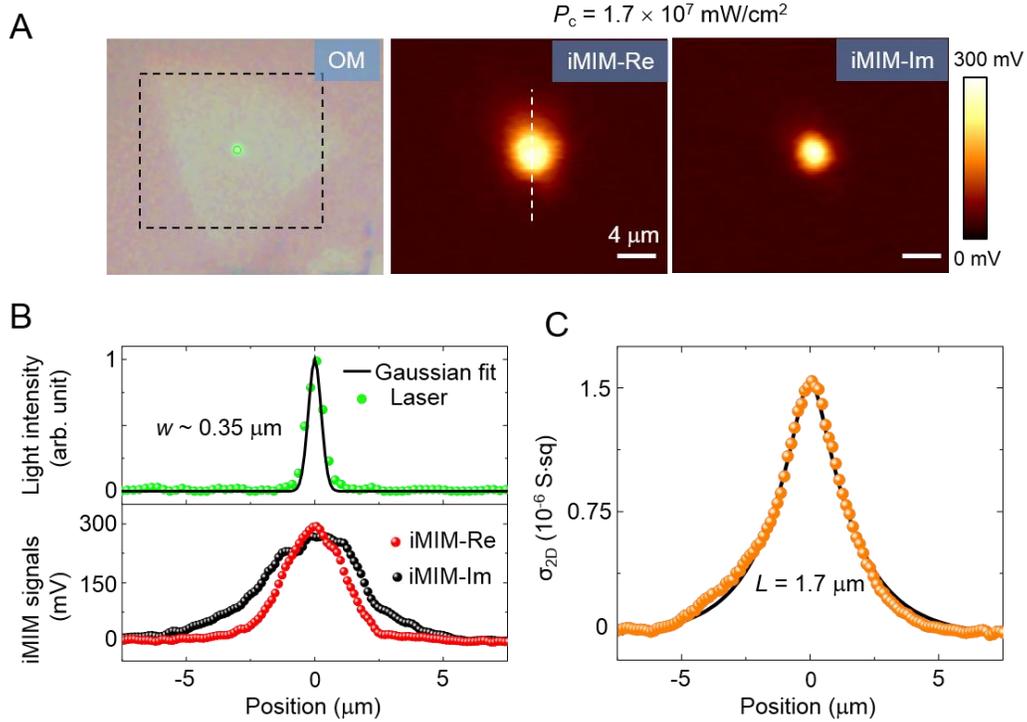

Figure S4. (*A*) Left to right: optical, iMIM-Re, and iMIM-Im images when a tightly focused laser beam illuminates the center of the WS$_2$ flake. The dashed box in the optical image shows where the iMIM data are taken. Both scale bars are 4 μm. (*B*) Line profiles of the laser spot (top) and iMIM signals (bottom) across the dashed line in *A*. (*C*) Measured photo-induced 2D conductance profile (orange dots) and numerical fit with a diffusion length of $L$ = 1.7 μm.

Figure S4 shows the diffusion analysis with a different focus of the laser beam, as seen in Fig. S4A. The Gaussian fit in Fig. S4B indicates that $w$ = 0.35 μm is near the diffraction limit. With this tight focus, there is some deviation from a Gaussian excitation spot, which explains the slightly oval-like shape in the iMIM diffusion image (Fig. S4A). Nevertheless, the numerical fit of the measured conductance profile to Eq. (2) in the main text still yields a diffusion length of $L$ ~ 1.7 μm (Fig. S4C). As expected, the experimentally determined diffusion length extracted from the iMIM profile is independent of the size of the laser spot.

## S5. Photoluminescence spectra of the two WS$_2$ samples.

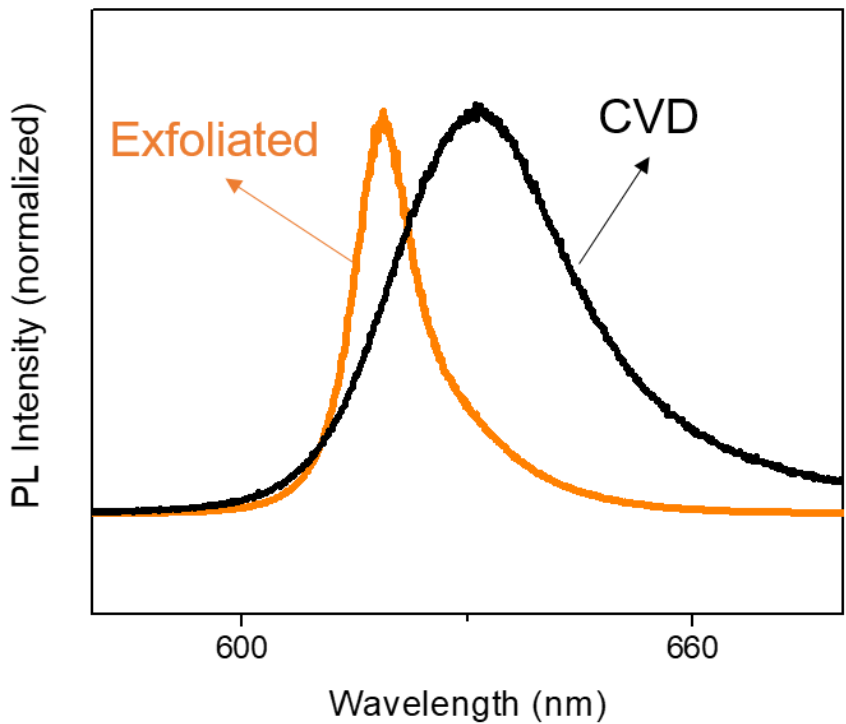

Figure S5. PL spectra of the two WS$_2$ samples in this work.

A comparison between typical photoluminescence (PL) spectra of the two WS$_2$ monolayers in this work is shown in Figure S5. It is clear that the PL linewidth is much narrower for the exfoliated Sample B with h-BN capping than that of the CVD-grown Sample A. It is generally accepted that the inhomogeneous broadening of the PL linewidth is strongly affected by atomic-scale defects [S7]. The results thus provide an indirect evidence on the much lower defect density in Sample B than Sample A.

## S6. Mobility measurement on the CVD-grown $WS_2$.

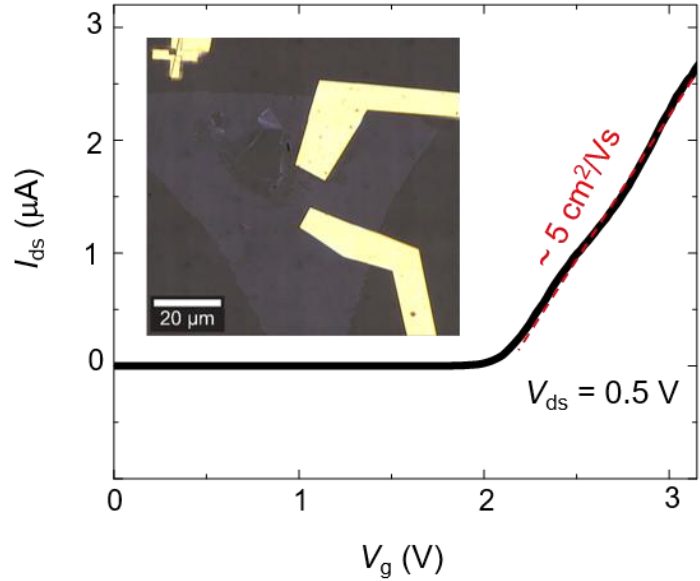

Figure S6. Mobility measurement of a CVD-grown $WS_2$ flake.

      Figure S6 shows the mobility measurement of a monolayer $WS_2$ flake grown by the same CVD process as the sample in the main text. Here the device was prepared by the standard E-beam lithography process. An ionic gel was used to gate the flake and n-type semiconductor behavior was observed. The capacitance of the ion gel is ~ $5 \times 10^{-6}$ F/cm$^2$. Because of the irregular electrode design and the hole in the flake, it is difficult to precisely define the channel length and width. We estimate the effective channel to be ~ 3 sq and the field-effect mobility is ~ 5 cm$^2$/Vs, which is consistent with the calculated value in the main text.

## S7. Quantitative analysis using the diffusion equation.

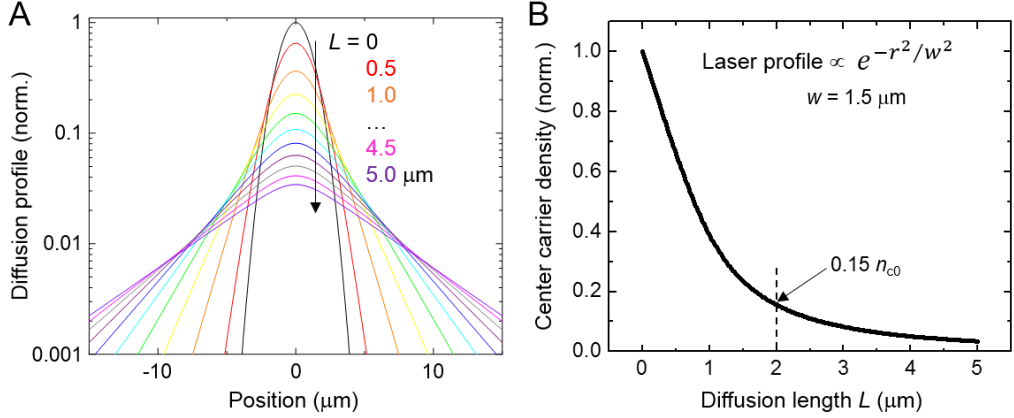

Figure S7. (***A***) Spatial distribution of charge carriers with a laser profile ~ exp($-r^2/w^2$) and different diffusion lengths *L*. (***B***) Carrier density at the center of the laser spot as a function of the diffusion length. The density at $L = 2$ μm is marked in the plot.

Using Eqs. (1) and (2) in the main text, we can quantitatively analyze the diffusion pattern to obtain parameters. Fig. S7A shows the spatial distribution of charge carriers with various diffusion lengths under an excitation profile $\propto e^{-r^2/w^2}$. Assuming $n_{c0} = \eta(P_c\tau/h\nu)$ is the density at the center of laser spot for $L = 0$ μm, the numerical solution indicates that $n_c = 0.15\, n_{c0}$ for $L = 2$ μm (Fig. S7B). Using the relation $\sigma_c = n_c q \mu$, one can calculate that $\eta \sim 0.1\%$, consistent with previous reports of IPCE in monolayer TMDs [S8, S9].

## S8. Density profile in the diffusion pattern.

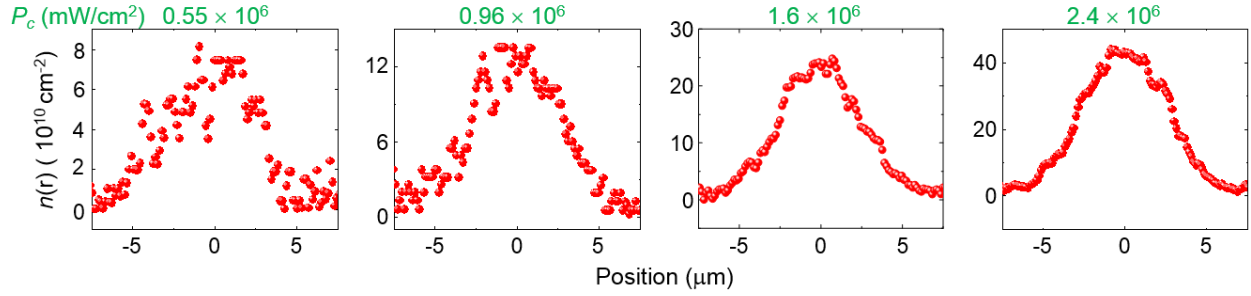

Figure S8. Calculated photo-carrier density profile for the diffusion patterns at various laser power.

Using Eq. (1) in the main text, where $L$ and $\tau$ are directly measured and $\eta$ estimated to be 0.1%, one can calculate the density profile in the diffusion pattern. Alternatively, using Eq. (3) in the main text, where $\sigma_{2D}$ is obtained from the iMIM data and $\mu \sim 4$ cm$^2$/Vs from the Einstein relation, we can also convert the photoconductivity at various excitation power to the carrier density, as shown in Fig. S8.